# Analysis of magnetic electron lens with secant hyperbolic field distribution

S. S. Pany[a*], Z. Ahmed[b], and B. P. Dubey[a]

  a. Electronics & Instrumentation Services Division, Bhabha Atomic Research Centre, Trombay, Mumbai-400085, India.
  b. Nuclear Physics Division, Bhabha Atomic Research Centre, Trombay, Mumbai-400085, India.

  E-mail: siddhartha.shankar.pany@gmail.com

ABSTRACT: Electron-optical imaging instruments like Scanning Electron Microscope (SEM) and Transmission Electron Microscope (TEM) use specially designed solenoid electromagnets for focusing of electron beam probe. Indicators of imaging performance of these instruments, like spatial resolution, have strong correlation with focal characteristics of the magnetic lenses which in turn have been shown to be functions of the spatial distribution of axial magnetic field generated by them. Owing to complicated design of practical lenses, empirical mathematical expressions are deemed convenient for use in physics based calculations of their focal properties. So, degree of closeness of such models to the actual field distribution determines accuracy of the calculations. Mathematical models proposed by Glaser[1] and Ramberg[1] have historically been put into extensive use. In this paper the authors discuss one such model with secant-hyperbolic type magnetic field distribution function, and present a comparison among these models, with results from finite element based field simulations as reference.

---

[*] Corresponding author

# Contents

## 1. Introduction

With improvement in computational capabilities of modern day computers and with better, more advanced field computation algorithms, modeling and simulation of complex electromagnetic problems are being possible with remarkable accuracy. Yet while it would be tempting to be convinced that simulations can eliminate the need for closed-form mathematical modeling of systems, two important questions should be considered:

**a.** Whether simulations can determine the degree of contribution of each input parameter to the output of a system

**b.** For a given desired output, whether simulations can determine the values of each contributing input parameter?

From the above questions it can be understood that the need for parameterization of a system remains pertinent despite having all resources. Whereas parameterization through computational EM simulations can a lengthy process of multiple solver runs with individual parametric sweep, even a crude mathematical model of the system can prove to be convenient in obtaining the rough estimates of the set of input parameters to a system, for a given desired output.

Focal characteristics of a solenoid electromagnetic lens have close dependency on its axial magnetic field distribution. In case of practical magnetic lens designs the variation of field intensity along the axis is, however a complicated function of the lens geometry and design, which cannot be derived easily considering the complexity of the lens structures. Therefore, it has been a regular practice to measure the field distribution experimentally and calculate the focal length. In order to design a lens for a given focal length and other parameters, it would necessitate the tedious labor of plotting the fields for a large number of cases. To avoid this exercise, attempts have been made to obtain empirical relations between the ampere-turns and focal lengths for various lens geometries. A better insight can be obtained if we can express the distribution of the field analytically to fit the curves which are obtained experimentally [1].

## 1.1 Choices of mathematical functions

In the vicinity of magnetic lenses, electrons in the beam experience Lorentz force due to distribution of magnetic field. Trajectory of electron beam in the form of radial position of the electrons as a function of axial position can be determined from solutions of equations of kinematics. In electron beam instruments like SEM and TEM, radial distances of electrons from the axis are of the order of a few nanometers i.e. negligibly small. It can be shown that the original equations of kinematics are reducible to simpler forms with linear powers of radial distance terms [1]. This is referred to as paraxial ray equation. Paraxial ray equation for solenoid magnetic lenses takes the following form:

$$\frac{d^2r}{dz^2} = -\frac{e}{8mV_r}B_z^2(z)r \qquad \ldots 1$$

where, r is the radial distance of the charged particle from the optical axis, e is the magnitude of charge, m is the mass, $V_r$ is the relativistically corrected acceleration voltage, and $B_z(z)$ is the axial distribution of the magnetic field interacting with the particle beam.

For parametric studies, a few fitting mathematical functions are generally used to represent the distribution of axial magnetic field. The most popularly used among them all is the Glaser [1] function, the expression for which is given as:

$$B_z(z) = \frac{B_0}{\left(1 + \frac{z^2}{a^2}\right)} \qquad \ldots 2$$

,where $B_0$ is the peak magnetic field and a is the half-width-at-half-maximum of the distribution. A similar distribution function by Ramberg [1] is given as:

$$B_z(z) = B_0 \operatorname{sech}^2\left(\frac{2.63z}{D}\right) \qquad \ldots 3$$

,where D is the bore diameter of the magnetic lens pole-pieces. The exponential model is yet another distribution function suitable to fit axial field distribution of solenoid magnetic lenses. Expression for the function is given as:

$$B_z(z) = B_0 e^{-\left(\frac{z^2}{a^2}\right)} \qquad \ldots 4$$

As far as analytical solution of the paraxial ray equation is concerned, it must be understood that it is of the form similar to Zero-Energy Schrödinger equation. In the equation, the expression for axial magnetic field distribution is analogous to the potential function in the Schrödinger equation and closed form analytical solutions can be obtained for such differential equations for only a handful of exactly solvable potential functions. Solutions obtained with the Glaser function are in simple trigonometric form and is thus the most widely used function in electron-optics since it makes computation of cardinal points of the lenses convenient.

However, the Glaser function has been shown to be accurate only for magnetic lenses with saturated pole-pieces [1]. Therefore modeling of practical lenses using this function might not be strictly speaking accurate, considering the fact that typical magnetic condenser lenses in SEMs are operated with pole-pieces way below magnetic saturation. The Ramberg model has been found to be more applicable to magnetic lenses having extremely small gap width between pole faces (S/D << 1) [1]. Out of all the existing models, the exponential model has been shown to approximate magnetic field distribution of practical condenser magnetic lens designs with greater accuracy [1]. Unfortunately the computation of the cardinal points in this case is difficult. [1]

In this paper, the authors discuss about another magnetic field distribution function which on one hand has been shown to fit that of a practical lens more accurately and on the other hand is in the form of an exactly solvable potential so analytical expression can be obtained for beam trajectory.

## 2. Analysis of secant-hyperbolic distribution function

### 2.1 Solution of paraxial ray equation

The expression for the secant-hyperbolic distribution function is:

$$B_z(z) = B_0 \operatorname{sech}(\alpha z) \qquad \ldots 5$$

,where $B_0$ represents the peak magnetic field at z=0 and $\alpha$ = 1.3170 / a, where a is the half-width at half-maximum of the distribution.

The paraxial ray equation for the function is:

$$\frac{d^2 r}{dz^2} = -\frac{e}{8mV_r} \frac{B_0^2}{cosh^2(\alpha z)} r \qquad \ldots 6$$

Equation-6 is in the form of one-dimensional Schrödinger equation for potential hole of modified Pösch-Teller type [2], which is of the form:

$$\frac{d^2 u}{dx^2} + \left\{ K^2 + \frac{\alpha^2 \lambda (\lambda - 1)}{(\cosh \alpha x)^2} \right\} u = 0 \qquad \ldots 7$$

Equation-6 can take the form of equation-7 where K=0 and:

$$\alpha^2 \lambda(\lambda - 1) = \frac{eB_0^2}{8mV_r} \qquad \text{... 8}$$

Values of λ can be obtained for the quadratic equation in equation-8 as:

$$\lambda = 0.5 \pm 0.5\sqrt{1 + \frac{eB_0^2}{2mV_r\alpha^2}} \qquad \text{... 9}$$

Solution of the Schrödinger equation in equation-7 can be obtained in terms of hypergeometric functions. General solution of the paraxial ray equation shown in equation-6 can be obtained incorporating the values of λ, which is given as:

$$r(z) = C_1 (\cosh \alpha z)^\lambda \, F_{1,2}\left[\frac{\lambda}{2},\frac{\lambda}{2};\frac{1}{2};-(\sinh \alpha z)^2\right]$$
$$+ C_2 (\cosh \alpha z)^\lambda (\sinh \alpha z) \, F_{1,2}\left[\frac{\lambda+1}{2},\frac{\lambda+1}{2};\frac{3}{2};-(\sinh \alpha z)^2\right] \qquad \text{... 10}$$

,where $F_{1,2}$ is an ordinary hypergeometric function and C1 and C2 are arbitrary constants.

For SEM condenser lenses, the boundary conditions in paraxial ray equations are specified as:

a. $r(z_0) = r_0$ (Beam diameter at the gun crossover)
b. b. $r'(z_0) = \left[\frac{dr}{dz}\right]_{z=z_0}$ (Divergence of the beam at the gun crossover)

Final values of constants C1 and C2 are obtained as solutions for the simultaneous equations-(a) and (b) and the final expression for trajectory of electron beam is obtained by incorporating the values of the constants in equation-10.

**2.2 Electromagnet design parameters**

**2.2.1 Peak Magnetic field:**

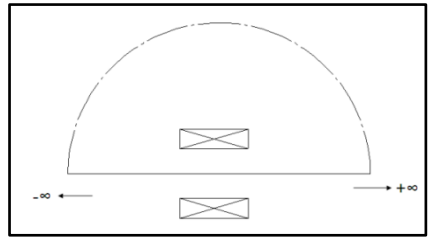

**Figure 1: Infinite length contour around solenoid electromagnet**

Peak magnetic field around the lens can be determined using Ampere's law. Let us consider an infinite semicircular contour C along the axis of the lens, as shown in figure-1. From Ampere's law we have:

$$\oint_C B.dl = \mu_0 NI \qquad \ldots 11$$

It can be observed that magnetic field along the semicircular arc is zero considering infinite boundary condition for the solenoid. Therefore, equation-11 can be simplified as:

$$\oint_C B.dl = \int_{-\infty}^{+\infty} B_z(z)dz = \mu_0 NI \qquad \ldots 12$$

Substituting values from equation-5 in equation-12, the value of $B_0$ can be given as:

$$B_0 = \frac{\mu_0 NI\alpha}{\pi} \qquad \ldots 13$$

### 2.2.2 Spread of magnetic field

The spread of axial magnetic field is measured as half-width-at half maximum (HWHM) of the field distribution. Let, *a* be the HWHM of axial field distribution with peak field $B_0$. Then, by definition, at $z=\pm a$, $B_z=0.5B0$. Substituting the values in equation-5, we have:

$$a = \frac{\text{sech}^{-1}(0.5)}{\alpha} = \frac{1.32}{\alpha} \qquad \ldots 14$$

Substituting the value of HWHM in equation-13, we have:

$$B_0 = \frac{1.32\mu_0 NI}{\pi a} \qquad \ldots 15$$

It is quite apparent that for a given value of NI and HWHM of the field distribution, the theoretical peak field for the secant hyperbolic field distribution turns out to be 1.32 times higher than that for the Glaser distribution in equation-2.

### 2.3 Focal characteristics

Computation of focal characteristics of the lens would involve determination of the axial positions of mid-focal plane, principal plane, focal plane and image plane. Mid-focal length is defined as the distance between the optical centre of the lens and the axial position at which radial distance of the ray becomes zero. If centre of the lens is at z=0 and $z_{mf}$ is the mid-focal length of the magnetic lens, then from equation-10 we have:

$$r(z_{mf}) = 0 \qquad \ldots 16$$

It can be observed that the expression in equation-10 is in the form of a linear combination of two hypergeometric functions. Finding generalized term for a zero for the expression would involve a functional term for an "inverse hypergeometric" function [3]. Considering the fact that no substantial mention of an inverse hyprgeometric function could be found in existing literature, the solutions of paraxial ray equation might not prove to be of help for obtaining generalized expressions for axial positions of cardinal planes of the lens. So, indirect methods have been used to determine focal properties of the distribution.

**2.4 Indirect methods for determination of focal characteristics**

There exist other approximate techniques for determination of optical characteristics of magnetic lenses. The Busch theorem [4] for focal length is one such approximate formula applicable to weak lenses more accurately. According to the theorem, the focal length *f* for a weak magnetic lens is given as:

$$\frac{1}{f} = \frac{e}{8mV_r} \int [B_z(z)]^2 \, dz \qquad \ldots 17$$

The condition for a magnetic lens to be considered as a weak lens is given as [4]:

$$\frac{V_r}{(NI)^2} > 0.009 \qquad \ldots 18$$

Applicability of equation-17 to the secant hyperbolic type magnetic field lens subject to the conditions in equation-18 can be ensured if:

$$\lambda < 1.03 \qquad \ldots 19$$

Focal length of a lens is defined as the distance between axial position of the principal plane and the plane at which an electron intersects the optical axis. With an analytical expression for the beam trajectory, mid-focal length can be obtained as solution of the equation r(z)=0 and with knowledge of the value of focal length, position of the principal plane would be possible to determine. It has been discussed in the previous chapter that obtaining a generalized expression for mid-focal length might not be possible owing to the infeasibility of obtaining a closed form expression for inverse hypergeometric function. However, for specific values of λ the hypergeometric function is reducible to simpler mathematical forms for which inverse functions can be formulated. For example, at λ=3 the first hypergeometric function assumes a form:

$$F_{1,2}\left[\frac{3}{2}, \frac{3}{2}; \frac{1}{2}; -(\sinh \alpha z)^2\right] = \frac{1 - 2(\sinh \alpha z)^2}{(\cosh \alpha z)^5} \qquad \ldots 20$$

The second term assumes a form:

$$F_{1,2}\left[2, 2; \frac{3}{2}; -(\sinh \alpha z)^2\right]$$
$$= 0.25(\operatorname{sech} \alpha z)^4 + \frac{(0.5 - (\sinh \alpha z)^2) \sin^{-1}(\sinh \alpha z)}{2 \sinh \alpha z \, (\cosh \alpha z)^5} \qquad \ldots 21$$

Substituting the expressions in equations-20 and 21 in equation-10 we have:

$$r(z) = (\text{sech}\,\alpha z)^2 \left[\left(2C1 + \frac{C2}{2}\sin^{-1}(\sinh \alpha z)\right)(0.5 - (\sinh \alpha z)^2) + C2(\sinh \alpha z)\right] \quad \ldots 22$$

It can be observed that, equation-22 mostly contains algebraic combinations of hyperbolic terms and finding solution for the inverse terms would be feasible.

## 3. Simulation & analysis of magnetic lens design

A prototype lens was designed to determine the coincidence of the existing mathematical models for magnetic field distribution with the same of magnetic lenses of practical design. Figure-2 shows the prototype ferromagnetic material shielded solenoid magnetic lens. It is a rotationally symmetric structure which consists of a solenoid electromagnet of ID=44mm, OD=84mm and has 1000 turns of copper wire of SWG 18. The solenoid is covered on all outer sides with magnetic return path which confine the magnetic field in a small space, between the pole-pieces. Material used for the magnetic structures is pure iron.

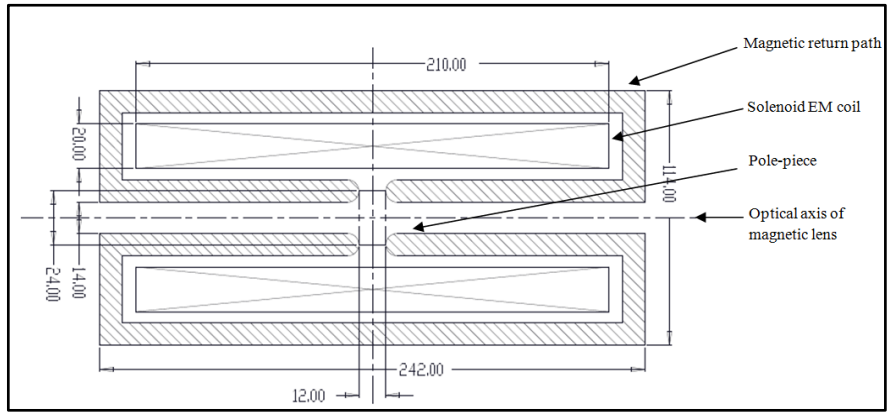

**Figure 2: Cross-sectional view of prototype magnetic lens**

Computational electromagnetic simulation of the assembly was carried out in a finite element based simulation software package called FEMM [5]. Excitation current to the solenoid was assigned to be 1A. The lens assembly being rotationally symmetric, one quarter section of it was simulated. Non-linearity in magnetization was taken into account during simulation of structures made of pure iron. As a benchmark guideline for FEM simulation, adjoining region with area of about 10 times that of the structure was considered as simulation space. Axial magnetic field was recorded at 1 million points along the optical axis of the structure. The peak magnetic field was found to be 0.0711 T with HWHM = 8.5mm and no part of the core was found to have gone into magnetic saturation.

Figure-3 shows the plots of axial magnetic field distributions corresponding to Glaser, Ramberg and the secant-hyperbolic functions and their comparison with the distribution obtained from simulation results. It can be observed that the spread of Glaser distribution is much higher than rest of the distributions and the Ramberg distribution has the least spread. The secant hyperbolic distribution seems to have a distribution closer to the distribution obtained from simulation.

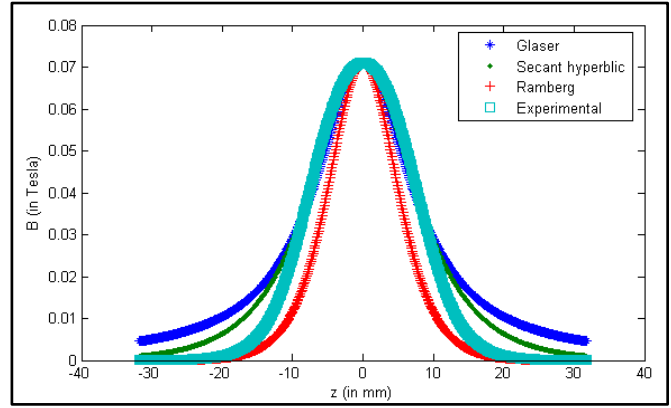

**Figure 3: Comparison among mathematical models and simulated result**

To compare between the distribution-functions, it is necessary to quantify their closeness with the simulation result. The factor that determines the accuracy of a mathematical distribution function is the value of focal length with respect to the value obtained from simulations. And it can be observed from equation-17 that, area under the squared axial magnetic field distribution curve is related to the focal length of a magnetic lens. Therefore, accuracy of a mathematical model can be accessed from the degree of closeness of the squared area of distribution curve with that of the simulated result.

A factor referred to as Closeness Quotient (CQ) has been considered for quantitative assessment of the degree of closeness of a model to the simulated distribution. CQ for a distribution is defined as:

$$(CQ)_{dist.} = \frac{|(B_{dist.})^2 - (B_{simulated})^2|}{(B_{simulated})^2} \qquad \ldots 23$$

Table-1 shows the CQ values for the three mathematical distribution functions.

| *Distribution* | *CQ* |
|---|---|
| Glaser model | 0.316 |
| Ramberg model | 0.2690 |
| Secant-hyperbolic | 0.1847 |

**Table 1: Comparison of CQ values for different mathematical models**

From the comparison it can be observed that the magnetic field distribution obtained from secant-hyperbolic model fits the one obtained from simulation of a practical lens design, better than the widely used models. Besides it can be observed that the Glaser distribution has much higher spread which might be the reason why it is considered to fit the situation of a lens whose magnetic core is beginning to saturate.

Simulations were repeated for magnetic lens designs with different pole shapes and coil sizes and positions.

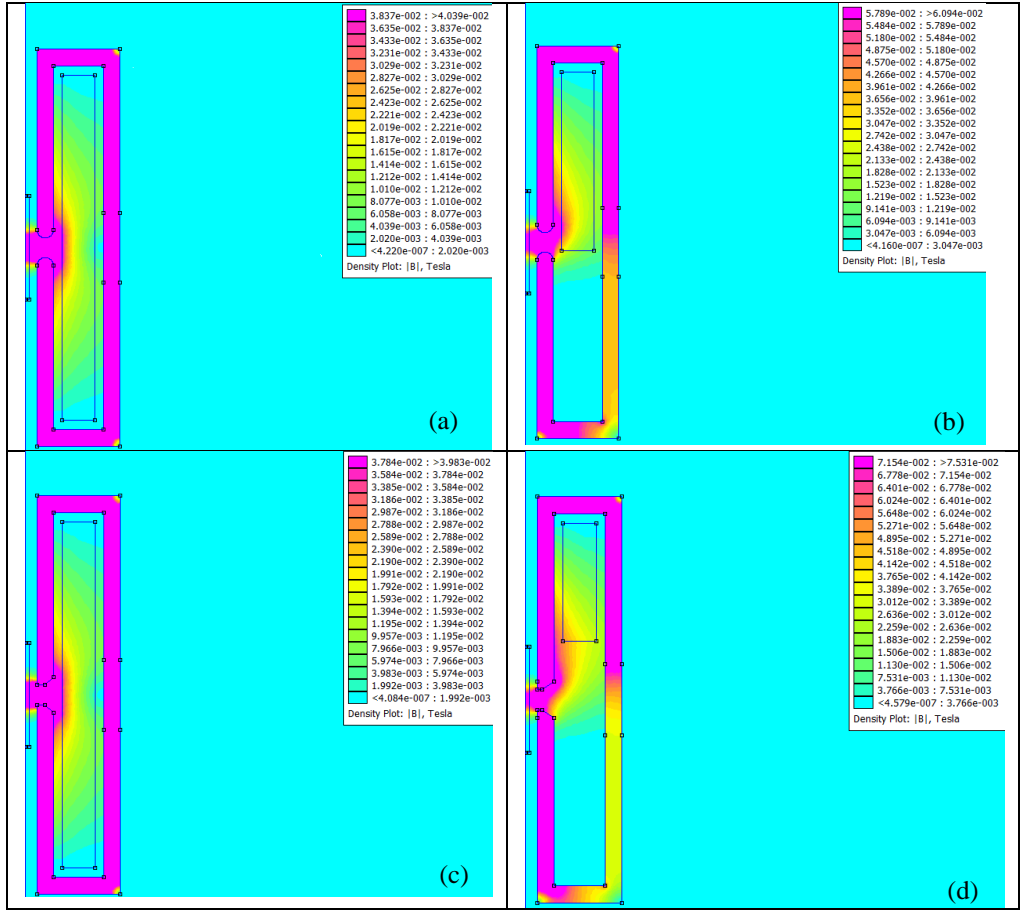

**Table 2: Magnetic field simulation results for: (a) Round pole-pieces, original coil size, (b) Round pole-pieces, reduced coil size, (c) Chamfered square pole-piece, original coil size, and (d) Chamfered square pole-piece, with reduced coil size**

Table-2 shows the spatial magnetic field distributions of two magnetic lenses with different pole-piece designs, obtained from simulation in FEMM. Simulation results for the widely used magnetic lenses with chamfered square pole-pieces have been shown in figures-(c) and (d). It was observed that, tangential magnetic field distribution along the optical axis of the lenses was not affected by the length of the solenoid coil, although the effect on magnetic field within the core of magnetic return path is quite visible. Possible explanation for the results could be attributed to the fact that, in the presence of non-saturating ferromagnetic core around the coil, field distribution near air-gap in the core is not significantly affected by the dimension of the solenoid coil, provided coil excitation is kept unchanged.

Simulations were also carried out by varying the S/D ratio [1] of the lenses. With increase in S/D ratio the HWHF of the field distribution curve was found to increase, however CQ value for the secant-hyperbolic function was still found to be less than that of the other two functions. Effects of magnetization characteristics of the core material were also studied by simulating the lenses with different grades of magnetic steel and they were found to have no significant influence on the axial field distribution as long as core saturation was avoided. In all the cases, CQ of the secant-hyperbolic function was found to be less than that of the other two functions.

## 4. Electron beam dynamics

Simulation of electron beam characteristics of the magnetic lenses were carried out to compare the results with that obtained from the secant-hyperbolic function. Typical energy of the beam in SEMs ranges from a few KeV to about 30 KeV and the size of the beam diameter at electron gun cross-over is about 50 micrometers, with a divergence of about 4 miliradian. The computations were carried out with these realistic values for beam parameters.

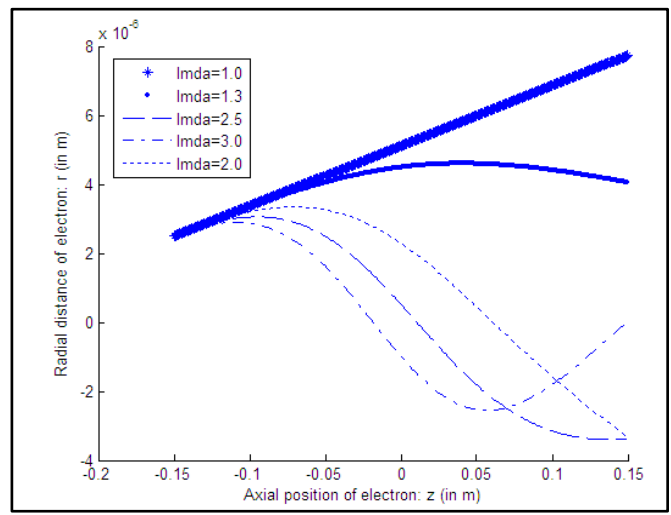

**Figure 4: Electron beam trajectory corresponding to different values of λ**

Figure-4 shows the trajectory of electron beam for different values of λ. It can be observed that for λ=1, the beam seems to move undeflected, and force on the beam keeps increasing for higher value of λ. Clearly the value of λ can be observed to be an indicator of focal strength of the lens.

From equation-17, the Busch expression for focal length for the secant-hyperbolic field distribution function is given as:

$$f = \frac{1}{2\alpha\lambda(\lambda - 1)} \qquad \ldots 24$$

From the expression it can be inferred that, the value of focal length as obtained for secant-hyperbolic function with a given spread, is a quadratic function of λ.

Considering the fact that the Busch expression for focal length is more appropriately applicable to weak lenses [4], the deviation between focal length values obtained from the formula and from simulation were recorded. Projective focal length [1] of the beam was obtained for different values of lambda using numerical methods from trajectory of horizontally emitted beams in the lens. Figure-5 shows the plot of focal lengths obtained from simulation and equation-24 against corresponding values of lambda.

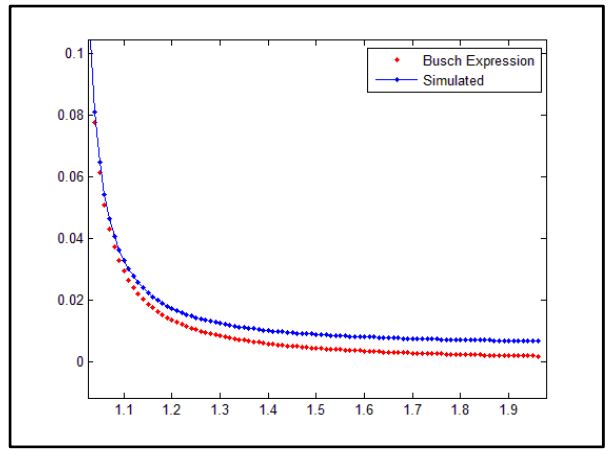

**Figure 5: Focal length (simulated and from Busch expression) v/s lambda**

The fact that the difference between the two focal length values seems to increase for higher values of λ, can be confirmed from Figure-6 which shows the plot of percentage difference between the focal lengths with the corresponding value of λ.

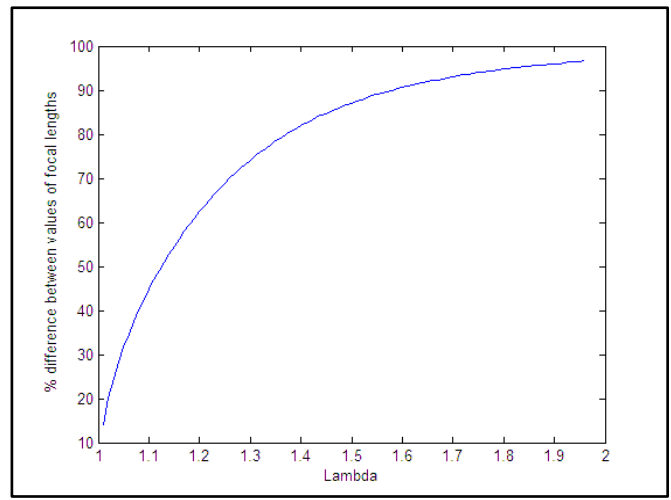

**Figure 6: Percentage difference in values of focal lengths in w.r.t λ**

It can be observed from the plot in figure-6 that although the difference between values of focal lengths at higher values of λ is significant, yet it seems to gradually saturate to a constant value about 100% beyond λ=2. It can be inferred that with a constant error for higher values of λ, the Busch expression for focal length can still prove to be useful which would make the use of results from the secant-hyperbolic function more convenient.

## 5. Conclusion

In this paper the difficulties encountered during solution of the paraxial ray equation for determination of electron beam trajectory in solenoid magnetic lenses were discussed with a purpose of bringing out the relevance of mathematical fitting functions for magnetic field distribution. The secant-hyperbolic function similar to the modified Pösch-Teller potential was presented in order to compare its accuracy with Glaser and Ramberg models. Magnetic lens characteristics were calculated using closed form solution of paraxial ray equation for the secant-hyperbolic function. The three distribution functions were compared with results obtained from FEM simulation of magnetic lens design models on the basis of their closeness quotient (CQ) values. By comparing the functions with simulation results for different designs and materials of lenses it was observed that the secant-hyperbolic function was fitting the field distribution of simulated lens better that the other two functions. Although the expressions for the general solution of the secant-hyperbolic function were found to be complicated for practical use, yet simpler forms could be obtained for some specific values of λ. Further simpler expressions for focal length were obtained with the Busch equation. Although mathematically the later is accurately applicable for weak magnetic lenses only, by comparison with simulated values it was observed that the error in focal length saturates to a constant value of 100% for higher strength of magnetic lenses, which makes it easier to use without compromising with accuracy of the value.

### Acknowledgments

S. S. Pany expresses his sincere gratitude to his colleagues and mentors for being a constant source of inspiration. He specially thanks Ms. Isha Gupta for her support and inspiration. He also thanks Mr. Shibaji Basu for his valuable comments and suggestions in this work.### References

[1] El-Kareh, A. B., El-Kareh, J. C.; *Electron Beam, Lenses, and Optics, Volume-1*; Academic Press; 1970

[2] Flügge, Siegfried; *Practical Quantum Mechanics*; Springer-Verlag Berling Heidelberg; 1994

[3] Senthilkumaran, M., Pandiaraja, D., Mayil Vaganan. *New Exact and Explicit Solutions of Generalized kdV Equations*. 2008. Appl. Math. comp. 209 693 – 699.

[4] Liebmann, G.; A unified representation of magnetic electron lens properties; 1955 Proc. Phys. Soc. B 68 737.

[5] D. C. Meeker, Finite Element Method Magnetics, Version 4.0.1 (03 Dec 2006 Build), http://www.femm.info– 12 –